# A large aperture reflective wave-plate for high-intensity short-pulse laser experiments

B. Aurand[1,2,3,4*], C. Rödel[1,5], H. Zhao[2,6], S. Kuschel[1,5], M. Wünsche[1,5], O. Jäckel[1,5],
M. Heyer[7], F. Wunderlich[7], M. C. Kaluza[1,5], G. G. Paulus[1,5] and T. Kuehl[1,2,3,4]

[1]*Helmholtz-Institut Jena; Helmholtzweg 4, 07743 Jena, Germany*
[2]*GSI Helmholtzzentrum für Schwerionenforschung GmbH; Planckstr. 1, 64291 Darmstadt, Germany*
[3]*Johannes Gutenberg University; Saarstr.21; 55128 Mainz, Germany*
[4]*EMMI ExtreMe Matter Institute; Planckstr.1; 64291 Darmstadt, Germany*
[5]*Institut für Optik und Quantenelektronik, Friedrich-Schiller University; Max-Wien-Platz 1, 07743 Jena, Germany*
[6]*Institut of Modern Physics; 509 Nanchang Rd.; 730000 Lanzhou, China*
[7]*Layertec GmbH; Ernst-Abbe-Weg 1; 9944 Mellingen, Germany*
[*]B.Aurand@gsi.de

Abstract: We report on a reflective wave-plate system utilizing phase-shifting mirrors (PSM) for a continuous variation of elliptical polarization without changing the beam position and direction. The scalability of multilayer optics to large apertures and the suitability for high-intensity broad-bandwidth laser beams make reflective wave-plates an ideal tool for experiments on relativistic laser-plasma interaction. Our measurements confirm the preservation of the pulse duration and spectrum when a 30-fs Ti:Sapphire laser beam passes the system.

## 1. Introduction:

Circularly polarized laser beams of high homogeneity are required for many applications. In Compton polarimetry, for example, the polarization purity of the scattered laser radiation is crucial to deduce the electron polarization [1]. When applied in laser-particle acceleration, circularly polarized lasers can significantly reduce the production of hot electrons and the associated high energetic radiation.

In the visible and near-infrared wavelength range, today's common technique to create circularly polarized light is the use of anisotropic transmissive elements which create, due to their birefringence, a phase shift between the electric field components parallel and perpendicular to the optical axis. Typical materials used for these retardation plates are mica- or quartz-crystals that are polished to the thickness

$$d_\Theta = \frac{\Delta\Theta \cdot \lambda}{\pi(n_{ao} - n_o)}; \quad d_{\lambda/4} = \frac{\lambda}{2(n_{ao} - n_o)} \qquad (1)$$

modulo integers of n·λ. Here, λ is the wavelength, ΔΘ is the desired phase shift and $n_{ao}$, ($n_o$) is the extraordinary (ordinary) index of refraction. Commercially available retardation plates are limited in size due to the risk of damage during polishing and the immense costs of growing large-size crystals. For the use of retardation plates with a significant bandwidth, typically two plates are bonded together in order to compensate the dispersion. This, however, increases the thickness of such optics and results in high-order dispersion and effects due to nonlinearities.

By using a well-designed multilayer mirror [2], an all-reflective approach to create a phase shift between the two polarization components of a light wave was proven to overcome these limitations. Utilizing magnetron or ion beam sputtering techniques, the multilayer design can be manufactured without practical restrictions of the physical size of the substrates. In addition and in contrast to large-aperture wave plates, the spatial beam quality will be preserved when high-quality substrates are used. Another advantage of reflective optics is that much less material is traversed by high-intensity radiation thus avoiding nonlinearities that alter the beam and pulse characteristics. Small residual effects on the pulse duration are determined by the multilayer coating design and the corresponding dispersion. Typically, PSMs with a group delay dispersion (GDD) of only 40 fs² can be designed and manufactured. As a result one obtains a reflective wave plate that features high polarization purity, high damage threshold and small aberrations due to high-order dispersion and nonlinear effects. The drawback of PSMs is that a fairly complicated geometry is required because the incident beam needs to have an equally intense s- and p-polarized field component on the mirror surface for creating circular polarization. Therefore, PSMs necessitate both a tilt and a rotation with respect to the incident polarization vector. As a consequence, the plane of incidence is tilted relative to the polarization plane and the output beam direction is oblique in space, specifically it encloses 45° with respect to the plane defined by the polarization direction and the input beam. This is not practical for most experimental environments.

In this paper we present a rotatable reflective wave-plate arrangement consisting of four mirrors that circumvents the problem and is capable of varying the ellipticity of the polarization without changing the beam direction.

## 2. Setup:

The assembly is an optical chicane consisting of four mirrors, each with an aperture of 120mm x 160mm supporting a beam diameter of 100mm. All mirrors have a broadband multilayer coating for the spectral range of λ=800±40 nm which is suitable for high-intensity Ti:Sapphire laser systems with pulse durations down to ≈25fs. In order to use the setup in a high-vacuum experimental environment, the mechanical parts of the assembly consist of aluminum, stainless steel, and vacuum compatible plastic components, see Fig.1. A balance weight allows a smooth rotation driven by a stepper motor via a tooth belt. The total weight of the apparatus is ≈25 kg, i.e. it can easily be implemented in a high-intensity laser experiment. The rotation axis is exactly in the center of the entrance and exit mirrors which enables a rotation of the assembly without a displacement of the laser beam. All mirrors are mounted at 45° with respect to the optical axis and can individually be adjusted by four screws such that the beam position and direction is invariant under the rotation of the device.

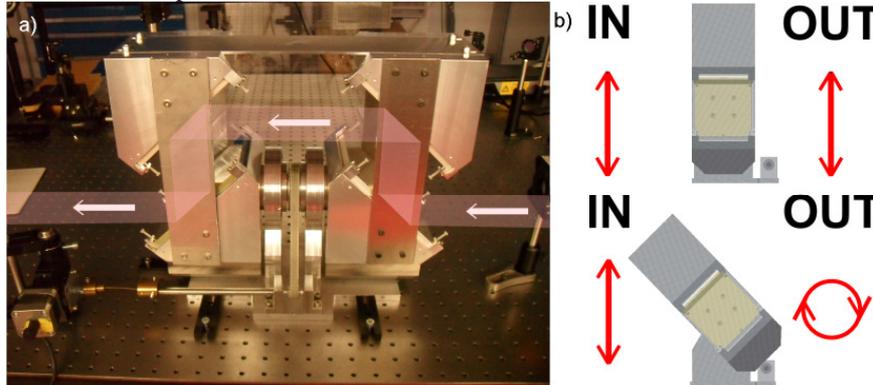

Fig. 1: a) Reflective wave-plate using a phase-shifting mirror for the generation of elliptically polarized high-intensity laser pulses. The beam experiences no displacement at the output when the assembly is rotated. b) Schematic of the functioning of the reflective wave-plate: When the polarization vector is parallel or perpendicular to the incidence plane of the four-mirror arrangement, the polarization at the output is unchanged (upper panel). A rotation of the assembly to 45° provides equal field components for s- and p-polarization on all four mirrors when the reflectivity is 100%. Thus a circularly polarized laser beam is generated when the accumulated phase difference between the s- and p-component is λ/4.

One of the four mirrors is a PSM providing a phase shift of 90°. The three other mirrors are designed to avoid any degradation of the ellipticity produced by this mirror. In the case of standard dielectric mirrors, the coating is typically intended for the reflection of pure s- or pure p-polarized light. Accordingly, a possible phase shift between the s- and p-component does not have any significance when they are used properly. However, such a phase shift is crucial if the mirrors are used for elliptically polarized light or in a tilted arrangement. As a consequence, standard high-reflecting multilayer coated mirrors would introduce an unknown and wavelength-dependent phase shift if used in our setup. Therefore we use three especially designed 0°-PSMs in our setup together with the 90°-PSM.

Similar to the 90°-PSM, the 0°-PSM layer design was calculated using the matrix-method. A reflectivity of more than 98% was achieved over the full bandwidth. The optimization routine is similar to the method described in [2] and the final design parameters are given in Fig. 2. It shows that a simple two-layer design using silver and a protective dielectric layer gives already reasonable results for a 0°-PSM. However, a higher reflectivity can be reached by using a more complex multilayer.

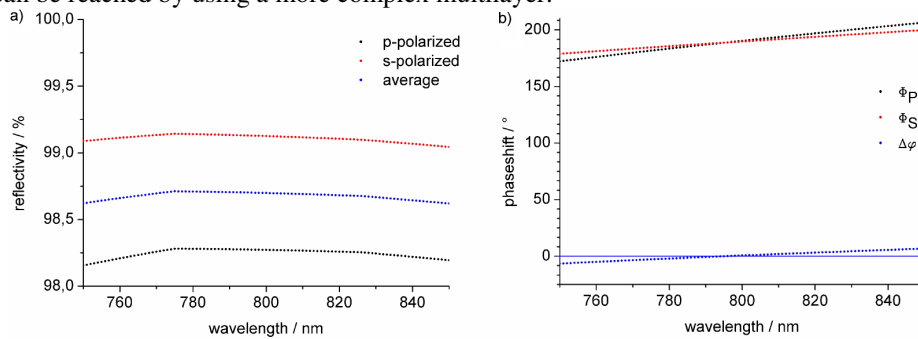

Fig. 2: Calculated reflectivity (a)) and phase (b)) for the design of the 0°-PSM. The reflectivity of each mirror exceeds 98% while no significant phase-shift between the s- and the p- polarized component is induced.

## 3. Experimental test of the apparatus:

The reflective wave-plate was tested using a Ti:Sapphire femtosecond laser system (FemtoPower Compact Pro) delivering 27-fs pulses with a bandwidth of λ=(790±50) nm at a 4kHz repetition rate. The beam size was limited by an iris to a diameter of 5mm.

The laser beam was linearly polarized perpendicular to the optical table. In order to enhance the polarization purity, we added a polarizing beam splitter (PBS) with an extinction ratio of larger than 1000:1. The residual ellipticity of the output beam was measured using a second PBS mounted on a rotation stage as an analyzer, as depicted in Fig.3a. For the measurement of the ellipticity the transmitted light was recorded with an energy-measurement head (Newport Model: 2936-C) while rotating the analyzer in steps of 10° for a full rotation. The transmitted energy P was averaged over a 5-s measurement period in order to achieve an accuracy in $\Delta P/P$ smaller than 1%. The ellipticity, which is proportional to the ratio of the minimum and the maximum of the transmitted electric field E over a full rotation of 360° was computed using:

$$\varepsilon = \frac{E_{min}}{E_{max}} = \sqrt{\frac{P_{min}}{P_{max}}} \qquad (2)$$

where $P_{min}$ and $P_{max}$ are the minimal and maximal measured energy. Accordingly, ε=0 denotes linearly polarized light and ε=1 circularly polarized light. When the two PBS are set up perpendicularly (crossed polarizer configuration) without using the polarizing mirror chicane, a value of ε=(1.7±0.2)% was found indicating an extinction ratio of the PBS of the order of 1000:1 or better. This value was assumed as the error of the ellipticity data for the following measurements with the reflective wave-plate.

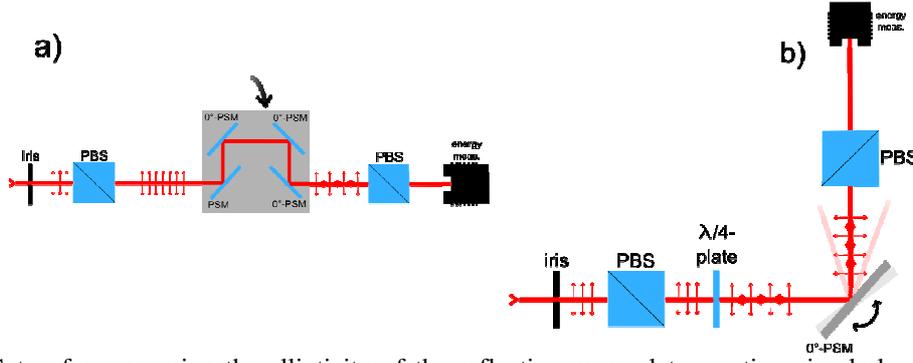

Fig. 3: a) Setup for measuring the ellipticity of the reflective wave-plate creating circularly polarized light (rotated to 45°) and linearly polarized light (rotated to 0°). (b) Setup for the measurement of the dependence of the phase shift of the 0°-PSMs on the angle of incidence. In this case the beam entering the PSM is circularly polarized by means of an additional λ/4 - plate.

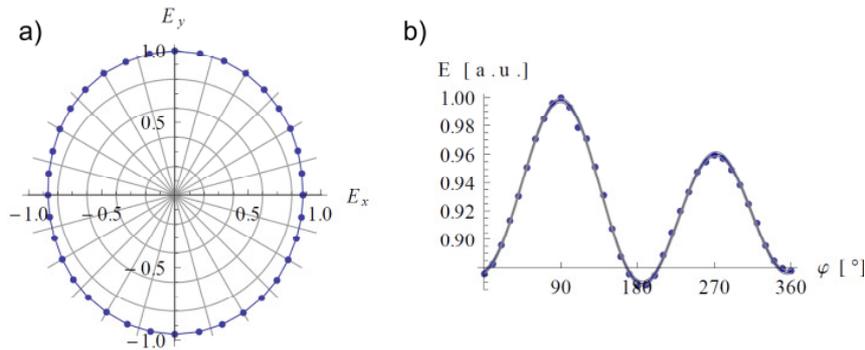

Fig. 4: Plot of the field distribution measured in case of a 45°-rotation of the reflective wave-plate, creating circular polarized light. The calculated ellipticity related to the fit is ε= (90.0±0.1)%.

The ellipticity of the reflective wave-plate was determined by fitting the following function to the data:

$$E(\varphi) = A \cdot \sin(2\varphi + \varphi_1) + B \cdot \sin(\varphi + \varphi_2) + C \qquad [3]$$

(blue line in Fig. 4). The amplitude of the first term (periodicity 2φ), describes the deviation from circular polarization. The second term, which can also be observed for circular polarization, has a periodicity of 1φ. It describes an artifact due to a beam deflection induced by the analyzing PBS.

We measured an ellipticity of the polarization induced by the reflective wave-plate of $\varepsilon_{RWP}$= (90.0±0.1)% for the case of a 45°-rotation angle and to $\varepsilon_{RWP}$= (2.1±0.02)% for 0° or 90° (Fig 4a, b). This maximum ellipticity is smaller than the ellipticity $\varepsilon_{PSM}$= (98.3±0.6)% [2] generated by a single PSM. This is attributed to an imperfect phase shift of the 0°-PSMs. It should be noted, however, that the ellipticity obtained with the reflective wave-plate is still higher than the ellipticity of $\varepsilon_{\lambda/4}$=75-85%, typically achieved with commercially available quarter wave-plates of the same size. Besides tolerances in the coating process, the purity of circular polarization is affected by the reflectivity of the p-component which is 1% smaller than the reflectivity of the s-component. This accumulates to ≈4% for all four mirrors and thus contributes ≈2% to the degradation of circular the polarization purity. The transmittance of the entire reflective wave-plate assembly is measured to be R= (91.9±0.4)%.

The achievable ellipticity is very sensitive to alignment. Therefore we measured the tolerance of the 0°-PSM by tilting the mirror and recording the ellipticity, see Fig. 3b. The result is shown in Fig 5. We found a tolerance of 2.5° centered around 43°. In this interval, the phase shift remains almost constant. However, the measured ellipticity is only $\varepsilon$=(94.5±0.5)%, see Fig.4. This means that the initial ellipticity drops by $\Delta\varepsilon$≈4% for this 0°-PSM. Accumulating this phase shift for three mirrors would result in a reduction of 12%. Fortunately, the combined phase shift of all four mirrors, results in a slightly better ellipticity of $\varepsilon_{RWP}$=90% as stated above. This is probably due to cancellations of the contributions of the individual mirrors.

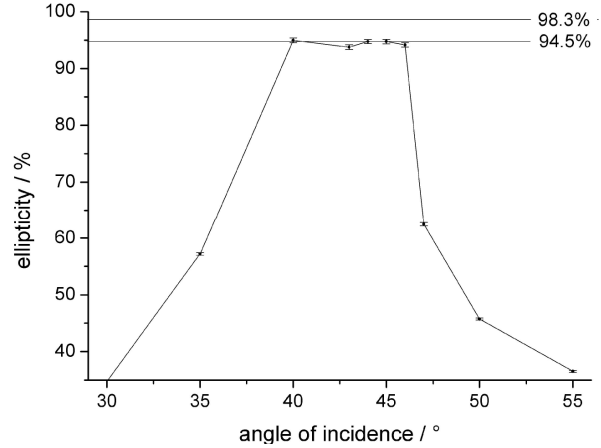

Fig. 5: Measured ellipticity of a circularly polarized laser beam ($\varepsilon$= (98.3±0.3)%) after passing a PSM designed to have a phase shift of 0°. When the angle of incidence is tilted from the design value of 45°, the ellipticity decreases. The stable regime to operate the PSM is $\Theta$=(42.5±2.5)°.

The spectrum of the laser pulses measured with a spectrometer after the reflective wave-plate (Ocean Optics, USB2000+), shows no bandwidth reduction or spectral interference as compared to the incident spectrum. In addition, the pulse duration was measured (APE SPIDER; 10-40fs) to be (27±1) fs, i.e. the incident pulse duration is preserved. Using the simulation results of the layer design, we expect a second-order dispersion $D_2$ of 40fs² for the 90°-PSM and less then 2fs² for the 0°-PSMs. Using the equation

$$\tau_{Final} = \tau_0 \sqrt{1 + \left(4\ln(2)\frac{D_2}{\tau_0^2}\right)^2} \qquad (4)$$

the corresponding pulse broadening can be calculated to be $\Delta t$=0.41fs which is smaller than the precision of the instrument. The damage threshold for the 0°-PSM was not measured separately but is calculated to be on the same order of magnitude than the one measured for the PSM [2], i.e. much higher than for a transmissive retardation plate [4].

**Summary and Outlook**

We have expanded the idea of a fully reflective wave-plate assembly to create circularly polarized light by a flexible device that can be easily implemented into an experiment with large-aperture high-intensity laser beams. The polarization can be adjusted between linear and circular polarization by rotating the device with a motorized stage. The maximum ellipticity of $\varepsilon_{RWP}$= 90%, is presently restricted by the imperfections of the coatings and the reflection losses of the p-polarized field component. However, the achieved ellipticity is clearly higher than the ones created with a typical large-aperture quarter-wave plate and can be increased with more complex multilayer coatings. The favorable attributes of PSMs like small GDD, high damage threshold, no influence on the transmitted spectrum, and the scalability to large aperture devices are transferred to a four-mirror optical setup that keeps the laser beam unchanged in position and direction. Therefore, the reflective wave-plate is an attractive tool for creating circularly polarized laser pulses in high-intensity laser experiments. A further step will be the generation of circularly polarized few-cycle laser beams using large broadband PSM and polarizing device.

**Acknowledgements**

B. Aurand acknowledges the support from the Helmholtz Association (HGS-Hire for FAIR) and the Helmholtz Institute Jena. C. Rödel appreciates funding from the Carl-Zeiss Stiftung. This work was supported in part by the Deutsche Forschungsgemeinschaft via project TR 18, Laserlab Europe and EMMI.